\documentclass[conference]{IEEEtran}

\usepackage{amsmath}
\usepackage{amsfonts}
\usepackage{amssymb}
\usepackage{epsf}
\usepackage{cite}
\usepackage{balance}
\usepackage{fancyhdr}
\usepackage{graphicx}
\usepackage{subfigure}
\usepackage[blocks]{authblk}
\usepackage[stable]{footmisc}
\usepackage{float}
\usepackage{fancyhdr}
\usepackage{gensymb}
\usepackage[bookmarks=false]{hyperref}
\usepackage{epstopdf}

\makeatletter
\def\ps@IEEEtitlepagestyle{%
  \def\@oddfoot{\mycopyrightnotice}%
  \def\@evenfoot{}%
}
\def\mycopyrightnotice{%
  {\footnotesize \copyright 2015 IEEE. Personal use of this material is permitted. Permission from IEEE must be obtained for all other uses, in any current or future media\hfill}
  \gdef\mycopyrightnotice{}
}

\newcounter{subeq}

\IEEEoverridecommandlockouts
\date{}
\begin{document}
\title{Indoor Positioning in High Speed OFDM Visible Light Communications}

\author{Mohammadreza Aminikashani}
\author {Wenjun Gu}
\author {Mohsen Kavehrad}

\affil{Department of Electrical Engineering\authorcr
The Pennsylvania State University, University Park, PA 16802\authorcr
Email: \{mza159, wzg112, mkavehrad\}@psu.edu\authorcr}

\maketitle
\begin{abstract}
Visible Light Communication (VLC) technology using light emitting diodes (LEDs) has been gaining increasing attention in recent years as it is appealing for a wide range of applications such as indoor positioning. Orthogonal frequency division multiplexing (OFDM) has been applied to indoor wireless optical communications in order to mitigate the effect of multipath distortion of the optical channel as well as increasing data rate. In this paper, we investigate the indoor positioning accuracy of optical based OFDM techniques used in VLC systems. A positioning algorithm based on power attenuation is used to estimate the receiver coordinates. We further calculate the positioning errors in all the locations of a room and compare them with those using single carrier modulation scheme, i.e., on-off keying (OOK) modulation. We demonstrate that OFDM positioning system outperforms its conventional counterpart.
 \end{abstract}

\section{Introduction}\label{INTRODUCTION}
\IEEEPARstart{V}{isible} light communication has been extensively studied recently as a promising complementary and/or alternative technology to its radio frequency (RF) counterparts particularly in indoor environments. VLC systems offer many attractive features such as higher bandwidth capacity, robustness to electromagnetic interference, excellent security and low cost deployment \cite{1.1,1.2,1.3,peng,1.4,1}. These systems however suffer from multipath distortion due to dispersion of the optical signal caused by reflections from various sources inside a room.

Orthogonal frequency division multiplexing (OFDM) has been proposed in the literature to combat intersymbol interference (ISI) caused by multipath reflections \cite{shieh2008coherent,gonzalez2005ofdm,elgala2009indoor,armstrong2006power,armstrong2009ofdm}. There have been several OFDM techniques for VLC systems using intensity-modulation direct-detection (IM/DD) such as DC-clipped OFDM \cite{kahn1997wireless}, asymmetrically clipped optical OFDM (ACO-OFDM) \cite{armstrong2006power} and PAM-modulated discrete multitone (PAM-DMT) \cite{lee2009pam}. In DC-clipped OFDM, a DC bias is added to the signal to make it positive. Hard-clipping on the negative signal peaks is used in order to reduce the DC bias required.  ACO-OFDM and PAM-DMT clip the entire negative excursion of the waveform. To avoid the impairment from clipping noise only odd subcarriers are modulated by information symbols in ACO-OFDM. In PAM-DMT, only the imaginary parts of the subcarriers are modulated such that clipping noise falls only on the real part of each subcarrier and becomes orthogonal to the desired signal.

In current indoor visible light positioning systems, several algorithms are proposed to calculate the receiver coordinates. In one approach, a photo diode (PD) is employed to detect received signal strength (RSS) information. The distance between transmitter and receiver is then estimated based on the power attenuation, and the receiver coordinates are calculated by lateration algorithm \cite{5,6}. In another approach, RSS information is pre-detected by a PD for each location and stored as fingerprint in the offline stage. By matching the stored fingerprints with the RSS feature of the current location, the receiver location is estimated in the online stage \cite{3}. In \cite{4}, proximity positioning concept was used relying on a grid of transmitters as reference points, each of which has a known coordinate. The mobile receiver is assigned the same coordinates as the reference point sending the strongest signal. Image sensor is another form of receiver which detects angle of arrival (AOA) information for the angulation algorithm used to calculate the receiver location \cite{2}. Other techniques were also proposed for the visible light system to improve the indoor positioning performance. In \cite{7}, Gaussian Mixture Sigma Point Particle filter technique was applied to increase the accuracy of the estimated coordinates. Accelerometer was applied in \cite{acc} such that the information on the receiver height is not required. To the best of our knowledge, the previous studies were built on the assumption of a low speed single carrier modulation or/and did not consider the multipath reflections. However, a practical VLC system would be likely to deploy the same configuration for both positioning and communication purposes where high speed data rates are desired. Furthermore, it was shown in \cite{gu2015} that multipath reflections can severely degrade the positioning accuracy especially in the corner and the edge areas of the room.

In this paper, to mitigate the multipath reflections as well as providing a high data rate transmission, we propose an OFDM VLC system that can be used for both indoor positioning and communications. The positioning algorithm employed is based on the RSS information detected by the PD and the lateration technique. We show that our proposed system can achieve an excellent accuracy even in dispersive optical channels.

The rest of the paper is organized as follows. In Section II, the system model and OFDM system configuration are briefly introduced. In Section III, the positioning algorithm is described. In Section IV, we present positioning accuracy of an indoor OFDM VLC system and compare it with its OOK counterpart. Finally, Section V concludes the paper.

\section{System Configuration}
\subsection{System Model}
\begin{figure}
\centering
\includegraphics[width = 8cm, height = 7.5cm]{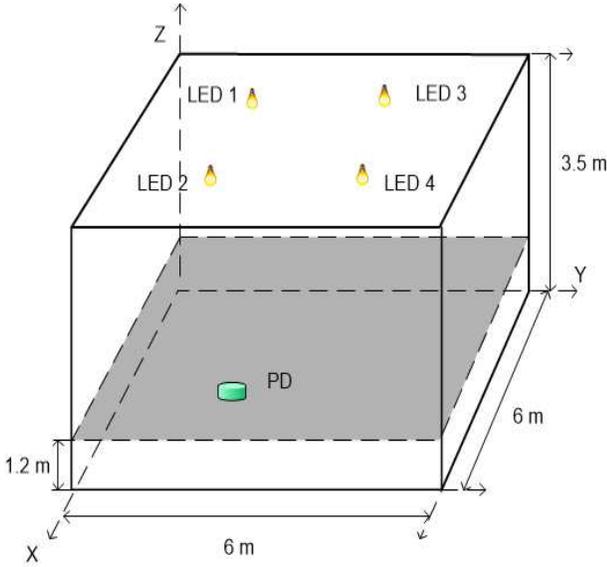}
\caption{System Configuration.}
\end{figure}
We consider a typical room shown in Fig. 1 with dimensions of 6 m $\times$ 6 m $\times$ 3.5 m where the LED bulbs are located at height of 3.3 m with a rectangular layout. Data are transmitted from these LED bulbs after they are modulated by the driver circuits. Each LED bulb has an identification (ID) denoting its coordinates which is included in the transmitted data. A PD as the receiver is located at the height of 1.2 m and has a field of view (FOV) of 70$^{\circ}$ and a receiving area of 1 cm$^2$. Strict time domain multiplexing is used where the entire OFDM frequency spectrum is assigned to a single LED transmitter for at least one OFDM symbol including a cyclic prefix (CP).
\subsection{OFDM transmitter and receiver}
Different OFDM techniques have been proposed for optical wireless communications in the literature. For the sake of brevity, ACO-OFDM is only considered in this paper as it utilizes a large dynamic range of LED and thus is more efficient in terms of optical power than systems using DC-biasing. However, the generalization to other techniques is straightforward. A block diagram of an ACO-OFDM communication and positioning system is depicted in Fig. 2. The data and LED ID code are combined as the input bits which is parsed into a set of $N/4$ complex data symbols denoted by $\mathbf{I}={{\left[ {{I}_{0}},{{I}_{1}},...{{I}_{{N}/{4}\;-1}} \right]}^{T}}$ where ${{\left( . \right)}^{T}}$ indicates the transpose of a vector. These symbols are drawn from constellations such as $M$-QAM or $M$-PSK where $M$ is the constellation size. For VLC systems using IM/DD, a real valued signal is required to modulate the LED intensity. Thus, ACO-OFDM subcarriers must have Hermitian symmetry. In ACO-OFDM, impairment from clipping noise is avoided by mapping the complex input symbols onto an $N\times 1$ vector $\mathbf{S}$ as
\begin{equation}
\mathbf{S}={{\left[ 0,{{I}_{0}},0,{{I}_{1}},...,0,{{I}_{N-1}},0,I_{N-1}^{*},0,...,I_{1}^{*},0,I_{0}^{*},0 \right]}^{T}}
\end{equation}
where ${{\left( . \right)}^{*}}$ denotes the complex conjugate of a vector. An $N$-point IFFT (inverse fast Fourier transform) is then applied creating the time domain signal $\mathbf{x}$. A CP is added to the real valued output signal to mitigate inter-carrier interference (ICI) as well as inter-block interference (IBI). All the negative values of the transmitted signal are clipped to zero to make it unipolar and suitable for optical transmission. This clipping operation does not affect the data-carrying subcarriers but decreases their amplitude to exactly a half. The clipped signal is then converted to analog and finally modulates the intensity of an LED.

At the receiver, the signal is detected by a PD and then converted back to a digital signal. The CP is removed and an $N$-point FFT is applied on the electrical OFDM signal. The training sequence is employed for synchronization and channel estimation as discussed in \cite{bilal}. A single tap equalizer is then used for each subcarrier to compensate for channel distortion and the transmitted symbols are recovered from the odd subcarriers and denoted by $\mathbf{\hat{I}}={{\left[ {{{\hat{I}}}_{0}},{{{\hat{I}}}_{1}},...{{{\hat{I}}}_{{N}/{4}\;-1}} \right]}^{T}}$. The LED ID is decoded and the transmitter coordinates are obtained which are fed to the positioning block along with the estimated channel DC gain as shown in Fig. 2. The receiver coordinates are estimated by employing the positioning algorithm detailed in the following section.
\begin{figure*}[t]
\centering
\includegraphics[width = 13cm, height = 6cm]{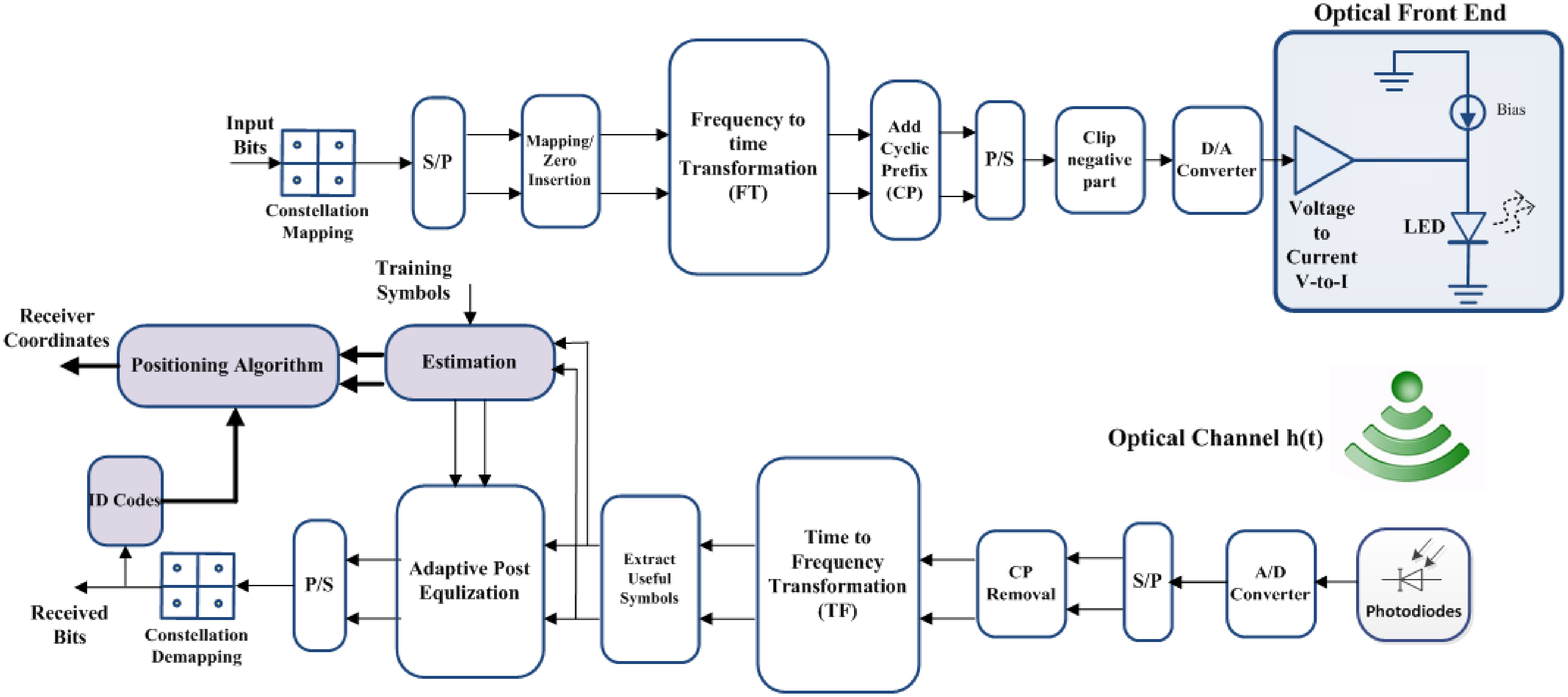}
\caption{OFDM transmitter and receiver configuration for both positioning and communication purposes.}
\end{figure*}
\section{POSITIONING ALGORITHM}
The distance between transmitter $k$ and receiver, $k=1,2,3$ and $4$, is estimated based on the estimated channel DC gain as \cite{zhang2014}
\begin{equation}\label{eq1}
{{d}_{k}}=\sqrt{\frac{\left( m+1 \right)A{{\cos }^{m}}\left( {{\phi }_{k}} \right){{T}_{s}}\left( {{\psi }_{k}} \right)g\left( {{\psi }_{k}} \right)\cos \left( {{\psi }_{k}} \right)}{2\pi {{{\bar{P}}}_{k}}}}
\end{equation}
where $A$ is the physical area of the detector, ${{\psi }_{k}}$ is the angle of incidence with respect to the receiver axis, ${{T}_{s}}\left( {{\psi }_{k}} \right)$ is the gain of optical filter, $g\left( {{\psi }_{k}} \right)$ is the concentrator gain, ${{\phi }_{k}}$ is the angle of irradiance with respect to the ${{k}^{th}}$ transmitter perpendicular axis, and $m$ is the Lambertian order. Assuming that both receiver and transmitter axes are perpendicular to the ceiling, ${{\phi }_{k}}$ and ${{\psi }_{k}}$ are equal and can be estimated as
\begin{equation}\label{eq2}
\cos \left( {{\psi }_{k}} \right)=\cos \left( {{\phi }_{k}} \right)=\left( H-h \right)/{{d}_{k}}
\end{equation}
where $H$ is the transmitter height and $h$ is the receiver height.

In (\ref{eq1}), ${{\bar{P}}_{k}}$ is the estimated channel DC gain obtained as
\begin{equation}\label{eq3}
{{\bar{P}}_{k}}=\frac{4}{N}\sum\limits_{i=1}^{{N}/{4}\;}{{{P}_{k,i}}}
\end{equation} 	
where ${{P}_{k,i}}$ is the power attenuation of the ${{i}^{th}}$ symbol transmitted from the ${{k}^{th}}$ transmitter and is calculated using the training symbols as
\begin{equation}\label{eq4}
{{P}_{k,i}}={{\left| \frac{{{{\hat{I}}}_{k,i}}}{{{I}_{k,i}}} \right|}}.
\end{equation}
For a compound parabolic concentrator (CPC), $g\left( {{\psi }_{k}} \right)$ is defined as
\begin{equation}\label{eq5}
g\left( \psi_{k}  \right)=\begin{cases}
\frac{{{n}^{2}}}{{{\sin }^{2}}\left( {{\Psi }_{c}} \right)}, & 0\le \psi_{k} \le {{\Psi }_{c}}  \\
0, & \psi_{k} >{{\Psi }_{c}}  \\
\end{cases},
\end{equation}
where $n$ and ${{\Psi }_{c}}$ denote the refractive index and the FOV of the concentrator, respectively. Considering (\ref{eq2}) and (\ref{eq5}),  (\ref{eq1}) can be rewritten as
\begin{equation}\label{eq7}
d_{k}^{m+3}=\frac{\left( m+1 \right)A{{T}_{s}}\left( {{\psi }_{k}} \right)g\left( {{\psi }_{k}} \right){{\left( H-h \right)}^{m+1}}}{2\pi {{{\bar{P}}}_{k}}}.
\end{equation}
Horizontal distance between the ${{k}^{th}}$ transmitter and the receiver can be estimated as
\begin{equation}\label{eq8}
{{r}_{k}}=\sqrt{{{d}_{k}}^{2}-{{\left( H-h \right)}^{2}}}.
\end{equation}
Then, according to the lateration algorithm, a set of four quadratic equations can be formed as follows
\begin{equation}\label{eq9}
 \begin{cases}
  {{\left( x-{{x}_{1}} \right)}^{2}}+{{\left( y-{{y}_{1}} \right)}^{2}}=r_{1}^{2}& \\
 {{\left( x-{{x}_{2}} \right)}^{2}}+{{\left( y-{{y}_{2}} \right)}^{2}}=r_{2}^{2}& \\
 {{\left( x-{{x}_{3}} \right)}^{2}}+{{\left( y-{{y}_{3}} \right)}^{2}}=r_{3}^{2}& \\
 {{\left( x-{{x}_{4}} \right)}^{2}}+{{\left( y-{{y}_{4}} \right)}^{2}}=r_{4}^{2}& \\
\end{cases},
\end{equation}
where $\left( x,y \right)$ is the receiver coordinates to be estimated and $\left( {{x}_{k}},{{y}_{k}} \right)$ is the ${{k}^{th}}$ transmitter coordinates obtained from the recovered LED ID.

By subtracting the first equation from the last three equations, we obtain
\begin{align}\nonumber
&\left( {{x}_{1}}-{{x}_{j}} \right)x+\left( {{y}_{1}}-{{y}_{j}} \right)y=\\\label{eq10}
&\left( r_{j}^{2}-r_{1}^{2}-x_{j}^{2}+x_{1}^{2}-y_{j}^{2}+y_{1}^{2} \right)/2
\end{align}
where $j=2,3$ and $4$. (\ref{eq10}) can be formed in a matrix format as $\mathbf{AX}=\mathbf{B}$ where $\mathbf{A}$, $\mathbf{B}$ and $\mathbf{X}$ are defined as
\begin{equation}
\mathbf{A}=\left[ \begin{matrix}
   {{x}_{2}}-{{x}_{1}} & {{y}_{2}}-{{y}_{1}}  \\
   {{x}_{3}}-{{x}_{1}} & {{y}_{3}}-{{y}_{1}}  \\
   {{x}_{4}}-{{x}_{1}} & {{y}_{4}}-{{y}_{1}}  \\
\end{matrix} \right],
\end{equation}
\begin{equation}
\mathbf{B}=\frac{1}{2}\left[ \begin{matrix}
   \left( r_{1}^{2}-r_{2}^{2} \right)+\left( x_{2}^{2}+y_{2}^{2} \right)-\left( x_{1}^{2}+y_{1}^{2} \right)  \\
   \left( r_{1}^{2}-r_{3}^{2} \right)+\left( x_{3}^{2}+y_{3}^{2} \right)-\left( x_{1}^{2}+y_{1}^{2} \right)  \\
   \left( r_{1}^{2}-r_{4}^{2} \right)+\left( x_{4}^{2}+y_{4}^{2} \right)-\left( x_{1}^{2}+y_{1}^{2} \right)  \\
\end{matrix} \right],
\end{equation}
\begin{equation}
\mathbf{X}={{[x\ y]}^{T}}.
\end{equation}
The estimated receiver coordinates can then be obtained by the linear least squares estimation approach as \cite{14}
\begin{equation}
\mathbf{\hat{X}}={{({{\mathbf{A}}^{\mathbf{T}}}\mathbf{A})}^{-1}}{{\mathbf{A}}^{\mathbf{T}}}\mathbf{B}.
\end{equation}
\section{SIMULATION AND ANALYSIS}
We assume an indoor optical multipath channel where transmitters and a receiver are placed in a room whose conﬁguration is summarized in Table I. The impulse response of the channel composed of the LOS and first three reflections is simulated with combined deterministic and modified Monte Carlo (CDMMC) method as discussed in \cite{12}.

We assume an OFDM system with a number of subcarriers of $N=512$ where the symbols are drawn from a 32-QAM modulation constellation. We assume a CP length of 16 and a data rate of 25 Mbps. Furthermore, to take LED nonlinearity into account, OPTEK, OVSPxBCR4 1-Watt white LED is considered in simulations whose optical and electrical characteristics are given in Table II. A polynomial order of five is used to realistically model the measured transfer function. Fig. 3 demonstrates the non-linear transfer characteristics of the LED from the data sheet and using the polynomial function. The four OPTEK LEDs are biased at 3.2V.
\begin{table}
\begin{center}
\begin{quote}
\caption{SYSTEM PARAMETERS}
\label{table2}
\end{quote}
\begin{tabular}{|c|c|}
  \hline
  \textbf{Room dimensions} &\textbf{Reflection coefficients} \\ \hline
  length: 6 m &    ${{\rho }_{wall}}$: 0.66\\
  width: 6 m  & ${{\rho }_{Ceiling}}$: 0.35\\
  height: 3.5 m & ${{\rho }_{Floor}}$: 0.60\\ \hline
 \textbf{Transmitters (Sources)} &\textbf{Receiver} \\ \hline
  Wavelength: 420 nm &Area $\left(A\right)$: 1$\times 10^{-4} \text{m}^{2}$\\
  Height $\left(H\right)$: 3.3 m &Height $\left(h\right)$: 1.2 m\\
 Lambertian mode $\left(m\right)$: 1 &Elevation: +90$^{\circ}$\\
 Elevation: -90$^{\circ}$ &Azimuth: 0$^{\circ}$\\
 Azimuth: 0$^{\circ}$ &FOV $\left({{\Psi }_{c}}\right)$: 70$^{\circ}$\\
 Coordinates: (2,2) (2,4) (4,2) (4,4) & \\
 Power for "1"/ "0": 5 W/3 W & \\ \hline
\end{tabular}
\end{center}
\end{table}
\begin{table}
\begin{center}
\begin{quote}
\caption{Optical and Electrical Characteristics of OPTEK, OVSPxBCR4 1-Watt white LED.}
\label{table2}
\end{quote}
{\small{
\begin{tabular}{|c|c|c|c|c|c|}
  \hline
  \textbf{Symbol} &\textbf{Parameter}  &\textbf{MIN} &\textbf{TYP} &\textbf{MAX} &\textbf{Units}\\ \hline
  $V_F$  &Forward Voltage &3.0 &3.5 &4 &$V$\\ \hline
  $\Phi$  &Luminous Flux &67 &90 &113 &lm\\ \hline
  ${{\Theta }^{1/2}}$  &50\% Power Angle &--- &120 &--- &deg\\ \hline
  \end{tabular}}}
\end{center}
\end{table}
\begin{figure} \centering
\subfigure[]{\includegraphics[width = 7cm, height = 5cm]{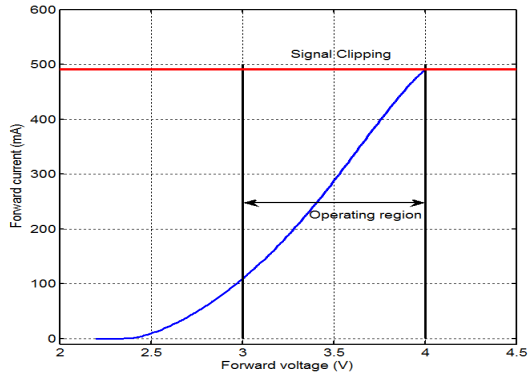}}
\subfigure[]{\includegraphics[width = 7cm, height = 5cm]{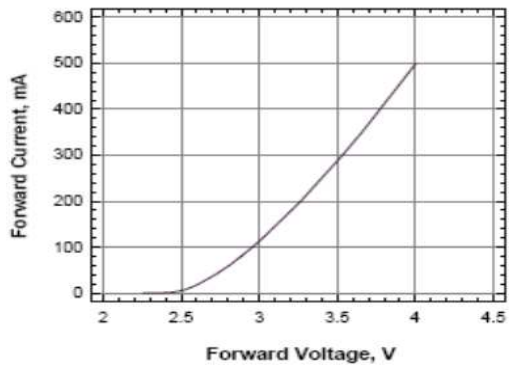}}
\caption{Transfer characteristics of OPTEK, OVSPxBCR4 1-Watt white LED. (a) Fifth-order polynomial fit to the data. (b) The curve from the data sheet.}
\label{led}
\end{figure}
Fig. 4 demonstrates the positioning error distribution over the room for the indoor OFDM VLC system under consideration. As it can be seen, the positioning errors are very small for the most locations inside the room, but become larger when the receiver approaches the corners and the edges due to the severity of the multipath reflections.
\begin{figure}
\centering
\includegraphics[width = 8cm, height = 7.5cm]{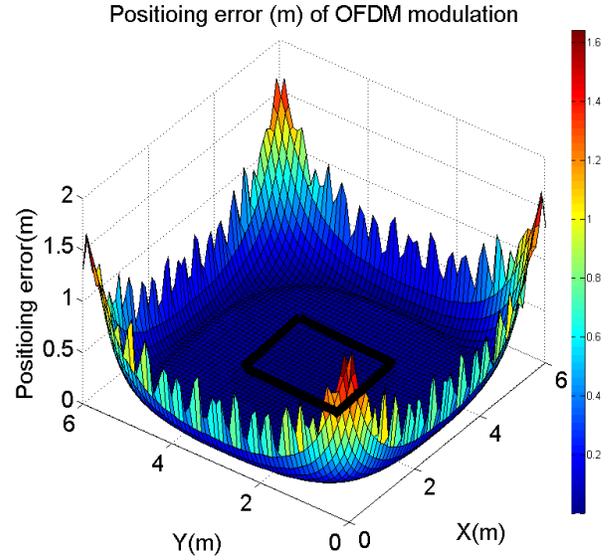}
\caption{Positioning error distribution for OFDM modulation.}
\end{figure}

Fig. 5, on the other hand, shows the positioning error distribution over the room for the indoor VLC system employing OOK modulation with the same data rate as that of the OFDM system. The positioning errors are relatively small within the rectangle shown in Fig. 5 where the LED bulbs are located right above its corners. However, the positioning error becomes significantly larger when the receiver moves toward the corners and the edges as the effect of the multipath reflections increases.
\begin{figure}
\centering
\includegraphics[width = 8cm, height = 7.5cm]{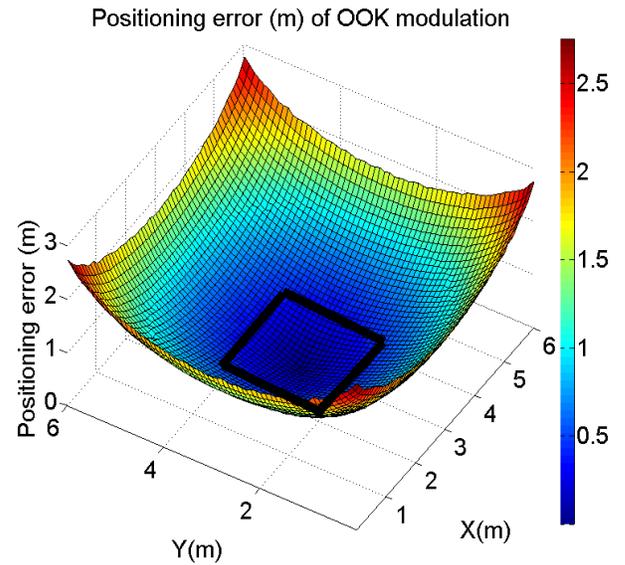}
\caption{Positioning error distribution for OOK modulation.}
\end{figure}

Figs. 6-7 present the histograms of the positioning errors for OFDM and OOK modulation schemes, respectively. For OFDM modulation, most of the positioning errors are less than 0.1 m and only a few of them are more than 1 m corresponding to the corner area. However, for OOK modulation, the positioning errors are widely spread from zero to around 2.3 m, and only a few of them are less than 0.1 m that correspond to the central area. From Figs. 6-7, it can be clearly seen that the OFDM system outperforms its OOK counterpart.
\begin{figure}
\centering
\includegraphics[width = 8cm, height = 7.5cm]{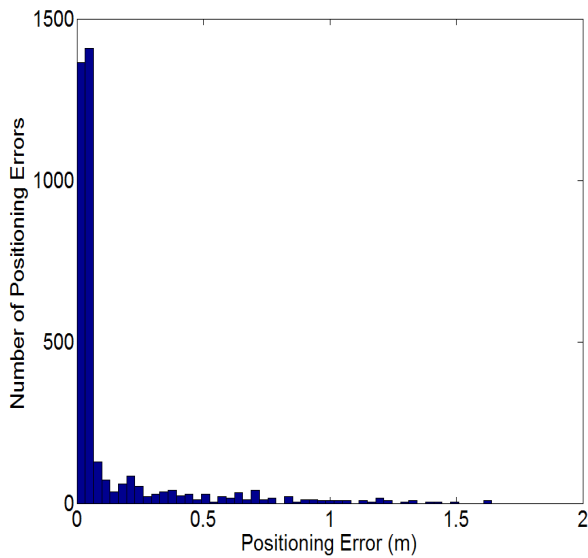}
\caption{Histogram of positioning error for OFDM modulation.}
\end{figure}
\begin{figure}
\centering
\includegraphics[width = 8cm, height = 7.5cm]{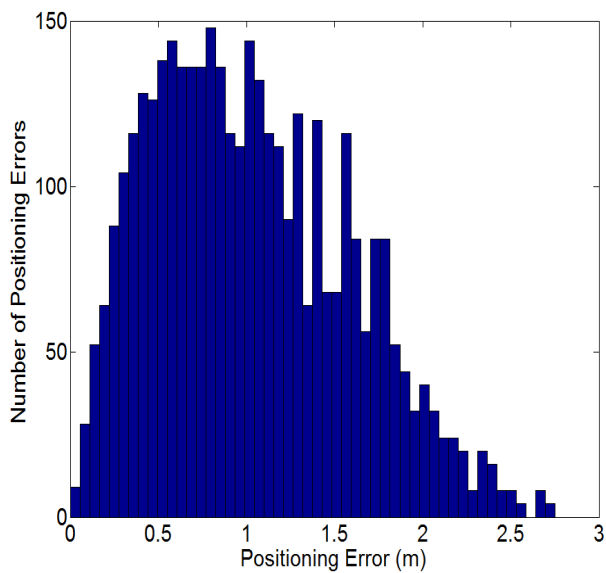}
\caption{Histogram of positioning error for OOK modulation.}
\end{figure}

Table III finally summarizes and compares the positioning errors of OFDM and OOK modulation schemes. OFDM modulation provides a much better positioning accuracy than OOK modulation for all the locations inside the room. Particularly, the root mean square (RMS) error is 0.04 m for the rectangular area covered perfectly by the four LED bulbs when OFDM modulation is used while it is 0.43 m for the OOK modulation. The total RMS errors are 0.53 m and 1.01 m for OFDM and OOK modulation schemes as the rectangular area covered by LED bulbs is only 11.1\% of the total area. The average positioning accuracy can be increased by optimizing the layout design of the LED bulbs in future.
\begin{table}
\begin{center}
\begin{quote}
\caption{Positioning error with/without reflections}
\label{table2}
\end{quote}
\begin{tabular}{|c|c|c|}
  \hline
  Positioning error (m) &\textbf{OFDM}   &\textbf{OOK}\\
  &\textbf{modulation (m)} & \textbf{modulation (m)}\\ \hline
  Corner (0, 0)  &1.95 &2.18 \\ \hline
  Edge (3 m, 0)  &0.72 &1.53 \\ \hline
  Center (3 m, 3 m)  &10$^{-6}$ &10$^{-5}$ \\ \hline
  RMS error of  &0.04 &0.43 \\
  the rectangular area & &\\\hline
  RMS error of   &0.53 &1.01 \\
  the whole room & & \\\hline
  \end{tabular}
\end{center}
\end{table}

\section{CONCLUSIONS}
In this paper, we have investigated and compared the positioning accuracy of IM/DD single- and multi-carrier modulation schemes for indoor VLC systems taking into account both nonlinear characteristics of LED and dispersive nature of optical wireless channel. Particularly, we have proposed an OFDM VLC system that can be used for both indoor positioning and communications. The training sequence used for synchronization has been adopted to estimate the channel DC gain. Lateration algorithm and the linear least squares estimation have been applied to calculate the receiver coordinates. We have shown the positioning error distribution over a typical room where the impulse response has been simulated employing CDMMC approach. Our results have demonstrated that the proposed OFDM system achieves an excellent accuracy and outperforms its OOK counterpart.
\section*{Acknowledgement}
The authors would like to thank the National Science Foundation (NSF) ECCS directorate for their support of this work under Award \# 1201636, as well as Award \# 1160924, on the NSF “Center on Optical Wireless Applications (COWA– \href{http://cowa.psu.edu}{http://cowa.psu.edu})”
\balance
\bibliographystyle{IEEEtran}
\bibliography{ref}

\end{document}